\documentclass[aps,pre,twocolumn,superscriptaddress,showpacs]{revtex4}

\usepackage{amssymb}
\usepackage{amsfonts}
\usepackage{rotating}
 \usepackage{amsmath}
\usepackage{graphicx}
\usepackage{epsfig}
\usepackage{graphics}
\usepackage{amsmath, amsthm, amssymb}

\newcommand{\lb}{\label}
\newcommand{\be}{\begin{equation}}
\newcommand{\ee}{\end{equation}}
\newcommand{\av}[1]{\langle #1 \rangle}
\newcommand{\bea}{\begin{eqnarray}}
\newcommand{\eea}{\end{eqnarray}}

\begin{document}

\title{Contagion dynamics in time-varying metapopulation networks}

\author{Su-Yu Liu}
\affiliation{State Key Lab. of Industrial Control Technology, Institute of Cyber-systems and Control, Zhejiang University, Hangzhou 310027, China}
\affiliation{Laboratory for the Modeling of Biological and Socio-technical Systems, Northeastern University, Boston MA 02115 USA}

\author{ Andrea Baronchelli}
\affiliation{Laboratory for the Modeling of Biological and Socio-technical Systems, Northeastern University, Boston MA 02115 USA}

\author{ Nicola Perra}
\affiliation{Laboratory for the Modeling of Biological and Socio-technical Systems, Northeastern University, Boston MA 02115 USA}

\date{\today} \widetext

\begin{abstract} 
The metapopulation framework is adopted in a wide array of disciplines 
to describe systems of well separated yet connected subpopulations. The subgroups/patches are often represented as nodes in a network whose links represent the migration routes among them. The connections has been so far mostly considered as static, but in general evolve in time. Here we address this case by investigating simple contagion processes on time-varying metapopulation networks. We focus on the SIR process, and determine analytically the mobility threshold for the onset of an epidemic spreading in the framework of activity-driven network models. We find profound differences from the case of static networks. The threshold is entirely described by the dynamical parameters defining the average number of instantaneously migrating individuals, and does not depend on the properties of the static network representation. Remarkably, the diffusion and contagion processes are slower in time-varying graphs than in their aggregated static counterparts, the mobility threshold been even two orders of magnitude 
larger in the first case. The presented results confirm the importance of considering the time-varying nature of complex networks.

\end{abstract}


\maketitle

Many natural and artificial networks evolve in time, with elements that appear, disappear and incessantly reshape their interaction patterns.  However,  the temporal dimension associated to their units and connections has started to become accessible only very recently \cite{holme11-1}, thanks to the increasing availability of empirical data \cite{Hui:2005,PhysRevE.71.046119,Onnela:2007,Gautreau:2009,10.1371/journal.pone.0011596,Tang:2010,Bajardi:2011,Stehle:2011nx,Miritello:2011,Karsai:2011}. Remarkably, the first analyses of dynamical processes unfolding on these empirical networks have shown that the full consideration of the time-varying patterns is responsible for a specific phenomenology, clearly distinct from the one observed on their time-aggregated counterparts. In particular, the temporal sequence of links and their concurrency play a crucial role when the time scale of the processes,  $\tau_{DP}$, is comparable to the time scale of the network, $\tau_G$
\cite{morris93-1,morris07-1,Rocha:2010,Isella:2011,Stehle:2011nx,Karsai:2011,Miritello:2011,dynnetkaski2011, albert2011sync,Parshani:2010,Bajardi:2011,pan11-1,consensus_temporal_nrets_2012,starnini_rw_temp_nets,perra12-1,perra12-2,ribeiro12-1}. This is the case, for example, of the spreading of sexual transmitted diseases in a population, ideas in scientific communities, and memes in social networks. 

However, the majority of the results obtained over the last years hold in two different limits. In the first case the network is considered as static. The time scale describing the evolution of its structure is much slower than the time scale of the dynamical processes, i.e.  $\tau_{G} \gg \tau_{DP}$. In the second case instead the networks is modeled as annealed. It evolves on a much faster timescale  i.e.  $\tau_{G} \ll \tau_{DP}$, and the dynamical processes effectively senses only time-averaged properties of the underlying network.
While these two limits are appropriate for such cases as the propagation of failures in technological networks (static network) \cite{havlin-book}, or to characterize the spreading of easily transmitted diseases, such as a novel pandemic (annealed networks) \cite{satorras01-1}, they are not appropriate for time-varying networks,  where $\tau_{DP} \sim \tau_G$ \cite{morris93-1,morris07-1,perra12-1,perra12-2}.

The study of dynamical processes in these time varying systems, often consists in introducing time windows of a certain size $\Delta t$. Thus, the connections between two consecutive intervals are integrated, generating many static networks. The dynamical processes are then let evolve on top of the sequence of these graphs. This procedure is extremely convenient when is possible to define a natural time window for the network, but it might introduce uncontrolled biases due to the size of the considered window \cite{perra12-1}.  For this reason, a different approach that fully addresses the time-varying nature of these graphs has been recently put forth, and has been adopted to characterize analytically the spreading of infectious diseases and random walks \cite{perra12-1,perra12-2}. It considers activity driven networks dynamics in which the nodes are characterized by an activity potential, defining their propensity towards creating links. The model reproduces important features found in many dynamical 
systems, being at the same time suitable for a simple analytical description of diffusive processes~\cite{perra12-1,perra12-2}. 

These efforts have considered systems in which nodes represent single individuals. Instead, here we address the case of time-varying metapopulation networks, in which nodes represent units/subpopulations with an internal dynamics coupled by the mobility of individuals. Indeed, each vertex contains an interacting subpopulation connected to its neighbors through links that represent routes of migration. Such reaction-diffusion processes have been extensively described in static metapopulation networks \cite{hanski97-1,tilman97-1,bascompte98-1,hanski04-1,colizza07-03,colizza07-5,colizza07-4,baronchelli2008bosonic}, but the understanding of the effects induced by dynamical connectivity patterns is still missing \cite{colizza07-5,balcan09-1,meloni11-1}. However, this is crucial to characterize phenomena as the movements of farmed animals, whose dynamics represents a major issue for public health, and where the evolutionary dynamics of the nodes (i.e., slaughterhouses, assembly centers
and also markets) and links has been shown to be fundamentally time dependent \cite{Bajardi:2011}.

We concentrate on contagion phenomena, having in mind elementary processes of information or disease spreading~\cite{barrat08-1}. For this reason, we consider the prototypical Susceptible-Infected-Removed (SIR) model \cite{kermac27-1,keeling08-1} for the internal dynamics. The diffusion of individuals is modeled by a simple homogeneous diffusion process characterized by a constant mobility rate $p$. We consider explicitly dynamical connectivity patterns between subpopulations describing them by activity driven models~\cite{perra12-1}. In this setting, we describe analytically the early phases of the reaction-diffusion process defining the necessary conditions for a global spreading  of the disease. We obtain the analytical expression for the threshold value of the mobility rate, $p^*$, that determines the outbreak of the spreading phenomenon. 
This finding generalizes the result concerning static networks, where the same threshold is determined by the first two moments of the degree distribution~\cite{colizza07-03,colizza07-5,colizza07-4}, and shows that also in the metapopulation framework the full consideration of the temporal patterns of the network is crucial for a satisfactory characterization of the observed phenomenology. In particular, we find that in time-varying metapopulation networks the threshold does not depend on the moments of the degree distribution, but rather is completely defined by dynamical observables such as the average number of instantaneously migrating individuals. Thus, the critical value of the mobility rate is independent of any time-aggregating representation of the network. We study and clarify the differences between time-varying metapopulation networks and their static counterparts by comparing the final number of diseased subpopulations. We observe the first being almost two order of magnitudes larger than the 
latter. The temporal nature of the connections in this case slows down the diffusion process requiring a much higher threshold. Our findings confirm series of results presented in recent literature~\cite{Rocha:2010,Isella:2011,Stehle:2011nx,Karsai:2011,Miritello:2011,dynnetkaski2011, albert2011sync,Parshani:2010,Bajardi:2011,pan11-1,consensus_temporal_nrets_2012,starnini_rw_temp_nets,perra12-1,perra12-2,ribeiro12-1} extending them for reaction-diffusion processes in the framework of metapopulation networks.

The structure of the paper is the following. In Section~\ref{meta_pop} we present the basic metapopulation framework. In Section~\ref{act} we describe activity driven models that we use to characterize the evolution of the links. In Section~\ref{meta_pop_act} within the metapopulation framework we analyze reaction-diffusion processes with time varying coupling patterns. In Section~\ref{conclu} we discuss the results and present the conclusions.

\section{Metapopulation framework}
\lb{meta_pop}

The metapopulation framework is used to describe the dynamics of systems characterized by well separated but connected units/subpopulations \cite{hanski97-1,tilman97-1,bascompte98-1,hanski04-1,colizza07-03,colizza07-5,colizza07-4,baronchelli2008bosonic}. Within this paradigm very different systems and processes can be described, such as chemical reactions, the evolution of populations, and the spreading of infectious diseases. In general, any spatially distributed set of subpopulations defined by a highly fragmented environment fits this description.

 Modeling such systems as networks is natural \cite{colizza07-03,colizza07-5,colizza07-4,baronchelli2008bosonic}. Nodes are the basics and isolated units, and links allow the movement of particles/individuals. Dynamical processes in metapopulation networks can be described as reaction-diffusion processes \cite{colizza07-4,baronchelli2008bosonic}. While the reaction refers to the dynamics taking place inside each node, the diffusion concerns the movement of individuals between connected nodes (see Figure~\ref{sk}-A for a schematic illustration). 
In this section we recall the mathematical formalism needed to tackle these problem for static networks.

\subsection{Reaction}

As mentioned above, we focus on disease/information spreading and contagion processes described using a SIR model \cite{kermac27-1,keeling08-1}. Individuals are divided in three different compartments according to their status, namely Susceptible (healthy, or unaware), Infected (and spreading), and Recovered (cured and no more infective). The evolution of the dynamics is defined by two different transitions. On the one hand,  susceptible individuals in contact with infected ones might become infected. This interaction happens with a rate $\beta$ and can be described as
\be
S+I\xrightarrow{\beta} 2I.
\ee
On the other hand, infected people recover from the disease spontaneously, with a rate $\mu$ described as:
\be
I\xrightarrow{\mu} R.
\ee
The inverse of this rate defines the characteristic infectious period, i.e. the average time each infected individual spends in the compartment $I$. 

The dynamics is determined by the value of these two parameters and by the interaction patterns among individuals. Here, we consider the simple case of homogeneous mixing. The nodes do not have an internal structure, and they all have the same probability of interaction. In particular, whether or not the disease will spread locally depends on the basic reproductive number  $R_0 = \beta/\mu$, describing the average number of secondary cases generated by an infected individual in a fully susceptible population \cite{kermac27-1,keeling08-1}. Diseases characterized by $R_0<1$ die out, i.e.,  the infected individuals are not able to sustain the spreading, while for $R_0 \ge1$ the disease is in general able to take off and spread locally.  However, the finite size of the subpopulations, and the stochastic nature of process, may lead to the extinction of the disease also in the latter case. In particular, the probability that a subpopulation experiences an outbreak give $n$ infected initial individuals is
\be
\lb{cond1}
P_{outbreak}=1-\frac{1}{R_0^n},
\ee
for all values of $R_0\ge1$ \cite{bailey75-1}. 

\subsection{Diffusion}

The diffusion of particles/individuals is a crucial component that drives the entire dynamics of the system. Many different diffusion mechanisms have been proposed with different levels of realism \cite{rvachev85-1,sattenspiel95-1,lloyd96-1,keeling00-1,keeling02-1,colizza07-5,balcan09-1,meloni11-1}. They all translate in the definition of a matrix $d_{ij}$  characterizing the probability that at each time step an individual from $i$ travels/diffuses to the subpopulation $j$. Here we consider a very simple diffusion mechanism with the following properties:
\begin{itemize}
\item at each time step the probability of traveling applies to all individuals in the node independently of their origins. These type of processes are Markovian diffusions. 
\item The probability $d_{ij}$ is set to be zero if there is not a link connecting the two subpopulations. 
\item Any individual in a node $i$, characterized by $k_i$ neighbors (degree) will have the same probability of reaching any of its neighbors: $d_{ij}=p\Delta t/k_i$ in a time interval $\Delta t$. The diffusion is homogeneous. The quantity $p$ defines the mobility rate of individuals.
\end{itemize}
In these settings, we consider a metapopulation network characterized by $V$ nodes and $N$ individuals. $N_i(t)$ is defined as the number of individuals inside node $i$ at time $t$. We assume that $\sum_i^{V}N_i(t)=N$ for any time $t$ i.e., the total population conserved. A convenient description of such systems is obtained considering degree classes, i.e., by assuming that nodes with the same number of neighbors, or degree, are statistically equivalent. 

The average number of individuals in a node of class $k$ is then:
\be
N_k=\frac{1}{V_k}\sum_{i|k_i=k}N_i,
\ee
where $V_k$ is the number of nodes of class $k$. The matrix $d_{ij}$ can be generalized for degree classes introducing $d_{kk'}$ as the probability that an individual diffuses from a node of class $k$ to a class $k'$ in an small interval $\Delta t$. Since we are considering homogeneous diffusion, we have $d_{kk'}=p \Delta t/k$. 

The variation in the number of individuals in a degree class $k$ in a interval $\Delta t$ can therefore be written by considering the master equation of a random walk on graph \cite{barrat08-1,newman10-1,noh04, baronchelli2010mean}:
\be
\lb{rand}
d_t N_k(t)= -p \Delta t N_k(t)+k\sum_{k'}N_{k'}(t)d_{k'k}P(k'|k).
\ee 
The first term represents the number of people leaving the class $k$. The last term considers the number of individuals arriving the class from all other connected nodes. In particular for uncorrelated networks, in which the probability that a link departing from a node of degree $k$ connects a node of degree $k'$ is independent  of  the destination, the stationary state of Eq.~\ref{rand} is:
\be
\lb{stationary_simple}
N_k=\frac{k}{\av{k}}\av{N},
\ee 
where $\av{N} = N/V$ is the average number of walkers per node. The equilibrium describes the final population size in each degree class. This results is well know  from the theory of dynamical processes on networks \cite{barrat08-1,newman10-1,noh04}. The linear dependence on $k$ remarks the effects of the heterogeneous connectivity patterns on diffusion processes.

\subsection{Invasion threshold}

In metapopulation networks the internal spreading of the disease ($R_0 \ge1$), does not necessarily imply a global spreading \cite{colizza07-03,colizza07-5,colizza07-4,baronchelli2008bosonic}. As schematically shown in Figure~\ref{sk}-B, in the limit $p=0$ the disease is not able to diffuse from one subpopulation to the other, but remains confined to the seeded subpopulations.  The spreading at the metapopulation level depends on the interplay between the time scale of the disease and the time scale of the diffusion. The mobility rate $p$ must be large enough to ensure the diffusion of infected individuals from seeded nodes to others, before the internal dynamics dies out. The minimal required value of the diffusion rate $p^*$ such that for $p>p^*$ the disease is able to spread in multiple nodes affecting a finite fraction of the system, defines the invasion threshold in metapopulation networks (see Figure~\ref{sk}-B). 

To identify the critical value $p^*$ for a general uncorrelated metapopulation network, let us consider a SIR epidemics started from a node of general degree $k$ and size $N_k$ characterized by $R_0 \ge 1$. As discussed in the previous section, above threshold the seeded node experiences an outbreak with probability given by Eq.~\ref{cond1}. In this case the number of infected individuals produced during the evolution of the epidemics can be written as $\alpha N_k$, where $\alpha$ depends on the details of the studied disease. In the limit $R_0 \to 1$ for a SIR model, $\alpha$ can be approximated by $\alpha \simeq \frac{2(R_0-1)}{R_0^2}$ \cite{murray2005-1}. In the SIR evolution, each infected individuals stays in the compartment $I$ for an average time $\mu^{-1}$ during which it can travel, and potentially infect other nodes. The diffusion matrix is $d_{kk'}$, so we can write the number of new seeds traveling from a node of class $k$ to a node of class $k'$ as:
\be
\lb{ll}
\lambda_{kk'}=d_{kk'}\frac{\alpha}{\mu}N_k.
\ee 
Let us define $D^{0}_k$ as the number of diseased subpopulation of degree $k$ at generation $0$. These nodes are experiencing an outbreak at the beginning of the process. They are the seeds of the infections. Nodes seeded by one of those in the first generation define the set $D^1_k$ at the following generation and so on. If the initial number of diseased nodes is small, the progression of the disease can be described using a tree-like approximation relating $D^n_k$ with $D^{n-1}_k$ \cite{ball97-1,harris89-1,vazquez06-2}. Following ref. \cite{colizza07-5} we can write:
\be
D_k^n=\sum_{k'} D_{k'}^{n-1}(k'-1)(1-R_0^{-\lambda_{k'k}})P(k|k')\left(1-\frac{D_k^{n-1}}{V_k} \right).
\ee
This equation considers that each node of degree $k'$ will seed the infection in $k'-1$ nodes (the total number minus the one from which the node got seeded). The conditional probability $P(k|k')$ measures the probability that each of the $k'$ neighbors subpopulations has degree $k$. The term $1-R_0^{-\lambda_{k'k}}$ considers the probability to observe an outbreak given the number of seeds $\lambda_{k'k}$ arriving from $k'$ to $k$. The last term in the sum describes the fraction of nodes in the degree $k$ still available for seeding. In the limit of $R_0 \sim 1$ and in the early stages of the epidemics we can simply the equation writing:
\be
D_k^n=\sum_{k'} D_{k'}^{n-1}(k'-1)\lambda_{k'k}(R_0-1)P(k|k').
\ee
For uncorrelated networks, homogeneous and Markovian diffusion dynamics the critical value of $p$ reads:
\be
p^*= \frac{\av{k}^2}{\av{k^2}-\av{k}}\frac{\mu}{(R_0-1)\alpha \av{N}}.
\ee
This result clearly shows the strong impact of the topological properties of the network on the global dynamics \cite{colizza07-03,colizza07-5,colizza07-4}. Indeed, for heavy-tailed networks the ratio $\frac{\av{k}^2}{\av{k^2}-\av{k}}$ is typically small, zero in infinite network size, implying that the heterogeneities of the metapopulation network favors the global spreading of epidemic disease. In this case the critical value of $p$ is much smaller than the respective value for homogeneous  networks. 

\begin{figure}
\begin{center} 
 \includegraphics  [width=0.4\textwidth] {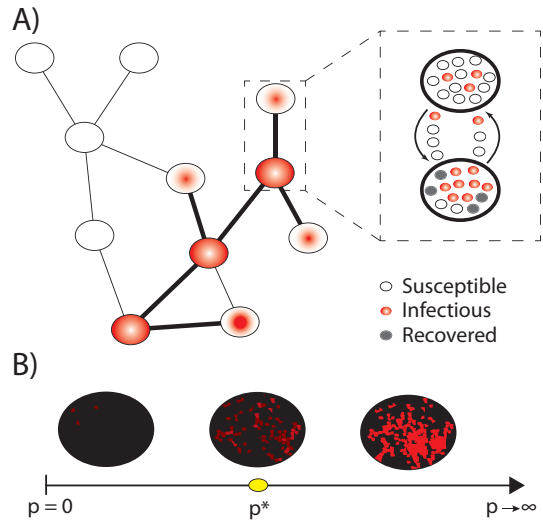}
\caption{(Color Online) Schematic representation of a metapopulation system and its global threshold. In panel A) we show, on the left, the diffusion process on the macroscopic level. Red (gray) nodes represent subpopulations experiencing a local outbreak. White nodes represent subpopulation not yet reached by the disease. Nodes with red (gray) center and white outer layer instead represent subpopulation that have been seeded by diseased population through the thick links connecting them. On the right we show the microscopic diffusion process. Each subpopulation contains individuals divided in three compartment according to their disease status. White are susceptible (white), red (light gray) infectious and dark gray recovered. The arrows represent the flows of migration coupling the internal dynamics. In panel B) we show the critical value of the mobility rate $p^*$. Below this point just a infinitesimal fraction of subpopulations is affected by the disease. For values equal or larger than the critical point the system reach a regime in which a finite fraction of 
subpopulations  is  affected by the spreading.  }
\lb{sk}
\end{center}
\end{figure}

\section{Activity driven models}
\lb{act}

In the following, we adopt the activity driven modeling of time varying networks, which represents a natural framework to study dynamical processes that evolve at the same time scale of the change of the networks \cite{perra12-1}. Each node is characterized by an activity potential $x$,  describing its propensity to establish links. 
Formally, $x$ is defined as the number of
interactions that a node performs in a characteristic time window of
given length $\Delta t$, divided by the total number of interactions
made by all nodes during the same time window. 
The distribution $F(x)$ defines the probability that randomly chosen node $i$ has activity
potential $x$, and the latter is bounded in the interval
$\epsilon \le x_i \le1$.
In the model, $N$ disconnected nodes are initially considered. To each one of them is assigned the activity/firing rate $a_i = \eta x_i$, defining the probability per
unit time to create new contacts or interactions with other
individuals. The rescaling factor $\eta$ is defined such that the average number of active nodes per unit time in the system is $\eta \av{x}N$. In this setting, the generative process is defined
according to the following rules:
\begin{itemize}
\item At each discrete time step $t$ the network $G_t$ starts with $N$
  disconnected vertices;
\item With probability $a_i \Delta t$ each vertex $i$ becomes active
  and generates $m$ links that are connected to $m$ other randomly
  selected vertices. Non-active nodes can still receive connections
  from other active vertices;
\item At the next time step $t + \Delta t$, all the edges in the
  network $G_t$ are deleted. 
 \end{itemize}
Thus, all interactions have a constant duration $\tau_G = \Delta t$, and nodes do
not have memory of any previous time steps. \\
If not specified otherwise, we will consider power-law distributions of activity potential i.e. $F(x)=Ax^{-\gamma}$ that reproduce behavior similar to what is observed in real data \cite{perra12-1}. At each time step the network is a simple random graph with low average connectivity. The
accumulation of connections that we observe by measuring activity
on increasingly larger time slices $T$ generates a skewed $P_T(k)$
degree distribution with a broad variability. 
Heterogeneities and nodes with a large number of connections (hubs) is
due to the wide variation of activity rates in the system and the
associated highly active agents. In activity driven models, the
creation of hubs results from the presence of nodes with high activity
rate, which are more willing to repeatedly engage in interactions.\\
The model allows for a simple analytical treatment \cite{perra12-1}. The average degree per unit time of a node $i$ characterized by an activity rate $a_i$ will be:
\be
\av{k_i}_t= m a_i + m\av{a}.
\ee
The first contribution is due to the $m$ links generated by the node when active. The second contribution is due to the links that reach node $i$ coming from other active nodes. At each time step the average degree per unit time in the network will be
\be
\av{k}_t=2m\av{a}.
\ee 
While a snapshot of the instantaneous network would show a set of
stars, the integrated network, defined as the union of all the instantaneous networks at
previous times, is not sparse. It is possible to show that the degree distribution
$P_T(k)$ of the integrated network at time $T$ takes the form:
\begin{equation}
P_T(k)\sim \frac{1}{T m \eta}F\left[ \frac{k}{T m \eta}\right], 
\end{equation}
in the limit of small $k/N$ and $k/T$ \cite{perra12-1}. Hence, 
the integrated degree distribution has the same function form of the activity potential. Fixing an opportune distribution for $F(x)$, the model reproduces the connectivity distribution observed in the data, and considers explicitly the dynamical evolution of its structure. 

\section{Metapopulation models with dynamical coupling patterns}
\lb{meta_pop_act}

In the next sections we present an extension of the framework and results obtained in static metapopulation networks, to systems characterized by time-varying topologies. We provide a first modeling approach for the study of relevant phenomena as the epidemic spreading in systems of farmed animals~\cite{Bajardi:2011}, or more in general any type of reaction-diffusion process characterized by a timescale comparable to the one describing the change in the topology of the metapopulation system.
 
In particular, we consider $V$ nodes and $N$ individuals in a metapopulation network. Each node is characterized by an activity potential $x$ extracted from a distribution $F(x)$. At each time step, individuals interact locally according to a SIR dynamics and move in or out a connected node by following a homogeneous and Markovian diffusion with rate $p$. 

 Interestingly, due to the effects of time varying coupling patterns~\cite{perra12-2}, we find expressions that are completely different from the ones observed in the static cases.

\subsection{Diffusion}

The diffusion mechanism has the same properties described in the previous sections, but now occurs on time-varying connections. At each time step $t$, each active node creates $m$ random links with randomly selected vertices, while inactive nodes can be selected by active ones. A networks $G_t$ is created and individuals can diffuse following its links with rate $p$. A convenient representation of such system is obtained considering activity classes. Considering the activity rate $a=\eta x$, the average number of population $N_a$ in nodes of activity class $a$ at time $t$ are:
\be
N_a(t)= \frac{1}{V_a}\sum_{i| a_i=a}N_i(t),
\ee
where $V_a$ is the number of nodes in activity class $a$. The variation of the number of individuals in each class of activity in an interval  $\Delta t $ is given by:
\begin{eqnarray}
\label{eq_prl_tot}
d_t N_a(t)&=&-ap \Delta tN_a(t)+a p m \Delta t \av{N}- \nonumber \\
&-&  p m \av{a } \Delta t N_a(t)+ \nonumber \\
&+&p \Delta t \int d a' a'  N_{a'}(t) F(a')
\end{eqnarray}
 The first two terms are contributions due to the activity of
the nodes in class $a$, while the final two terms represent the contribution to
inactive nodes due to the activity of the nodes in all the other
classes. In particular, the first term considers that active nodes will let a fraction $p\Delta t$ move in others nodes. The second term considers that active nodes will connected with $m$ others subpopulations getting a fraction $p\Delta t$ of their individuals. The third term takes into account that active nodes might connected with nodes in class $a$ getting a fraction $p\Delta t$ of their individuals. The last term considers the number of individuals arriving in nodes of class $a$ once these are selected by other active nodes.

 As discussed in more details in ref \cite{perra12-2}, the stationary state of the process is defined by the
infinite time limit $\lim_{t\to\infty}\dot{N}_a\left(t\right)=0$ leading to:
\begin{equation}
\label{beauty}
N_a=\frac{am\langle N\rangle+\phi_1}{a+m\langle a\rangle},
\end{equation}
 where the constant term $\phi_1=\int da aF(a)N_a$ is the average number of individuals potentially moving out from active nodes per
unit time. In order to better understand the physical meaning of this quantity let us consider Eq.~\ref{eq_prl_tot}. This can be rewritten as:
\be
d_t N_a(t)= -F_{out,a}(t)+F_{in,a}(t),
\ee
where the two quantities represent the out and in flows (negative and positive terms in Eq.~\ref{eq_prl_tot}) from nodes in the activity class $a$. Since the number of individuals in the system is conserved, summing over all activity classes, we can write:
\bea
 d_t \int  da N_a(t) VF(a)&=&- \int  da \;\ F_{out,a}(t)VF(a)+ \nonumber\\ 
 &+& \int  da \;\ F_{in,a}(t)VF(a)=0.
\eea
The two total contributions must be equal with opposite signs. Let us focus on the negative term. The integral over all the classes gives the total number of individuals moving out of any node in the system:
\bea
F_{out}(t) &=& \int da \;\ F_{out,a}(t) V F(a) \nonumber \\
&=&Vp\Delta t \int da \;\ N_a(t) (a+ m \langle a \rangle ) F(a),
\eea
We can then write:
\bea
F_{out}(t)&=& V p \Delta t\int da \;\ N_a(t) a F(a)+\\ \nonumber
&+& V p \Delta t m\int da \;\ N_a(t) F(a) \langle a \rangle,
\eea
so that:
\be
F_{out}(t)=V p  \Delta t\phi_1 + mVp \Delta t \langle a \rangle \langle W \rangle.
\ee
The first term defines the number of individuals moving out from active nodes in the system. The quantity $\phi_1$ is defined as average number of people potentially moving out of active nodes per unit time. It becomes the actual average in the limit $p=1$. The second term defines the number of individuals moving out from not active nodes that have been selected by active ones. 
 
The stationary state found in Eq.~\ref{beauty} differs substantially from the ones found in static ones  (Eq.~\ref{stationary_simple}), where the number of individuals in each node of degree $k$ is a linear function of the degree. This is not the case in time-varying networks, in fact. The difference is due to the effects of the dynamical connectivity patterns that play a crucial role in the characterization of the stationary state. In particular, in activity driven networks there is a saturation effect for large values of activities. Nodes with high activity have on average $k\sim m$ connections at each time step, and henceforth a limited capacity of collecting individuals \cite{perra12-2}.

\begin{figure}
\begin{center} 
 \includegraphics  [width=0.35\textwidth] {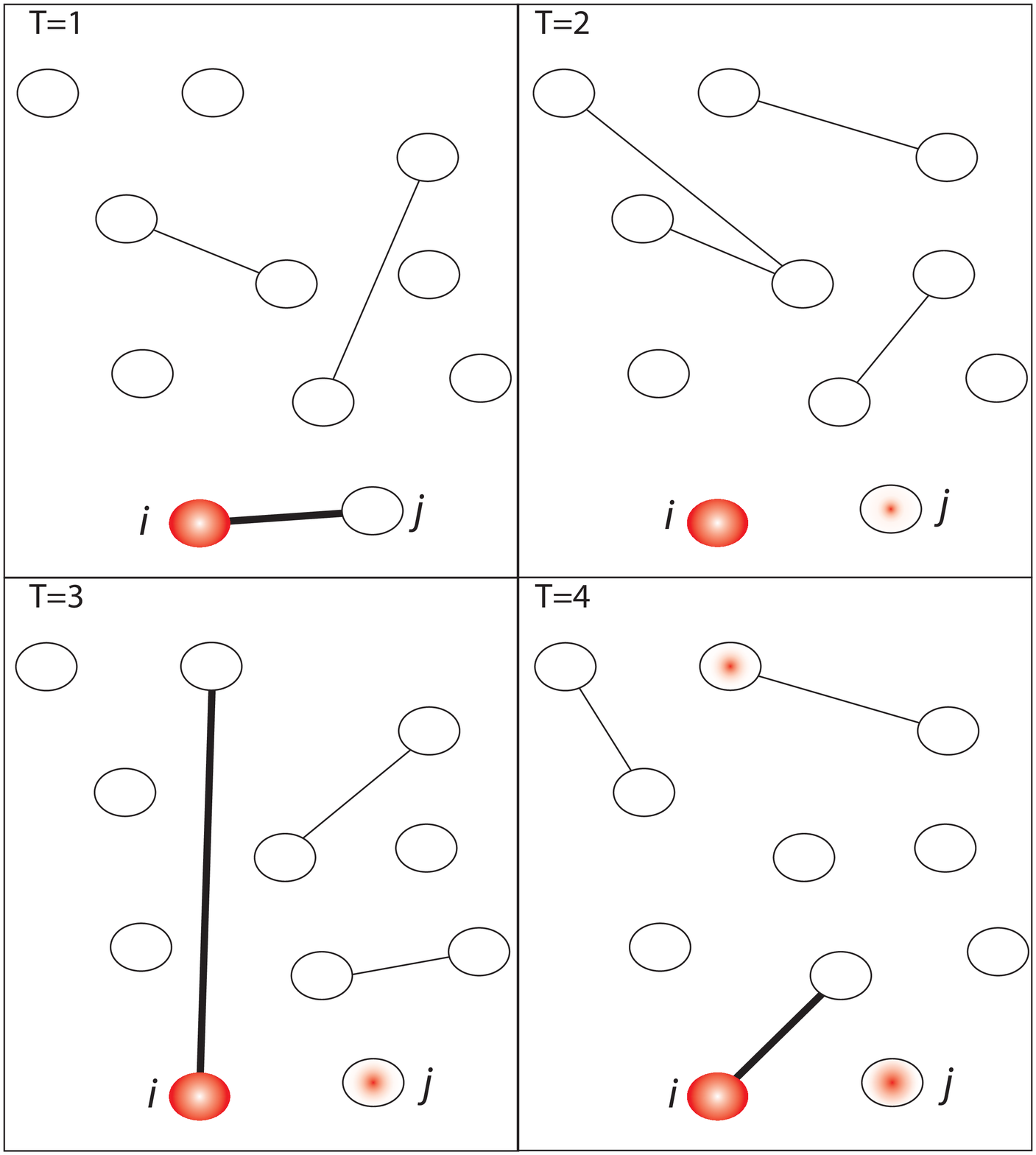}
\caption{(Color Online) Schematic representation of the macroscopic dynamics in time-varying metapopulation networks. In the first time step, $T=1$, just the subpopulation $i$ is diseased (red/gray node). It is the source of the infection. By chance it becomes active and creates $m=1$ random links. Through this connection, thick line, some infected individuals to move to the subpopulation $j$ that will starts to experience the disease at the next time step. Next, for $T=2$, three others nodes are active. All the selected nodes are not diseased so no infected individuals diffuse. At the time step, $T=3$, node $i$ is active again infecting another node. The process is repeated for all the other time steps,  driven by the activity of the each node. }
\lb{sk_activity}
\end{center}
\end{figure}

\subsection{Invasion Threshold}

To better illustrate the results, we focus our attention first on the case $m=1$, where each active node creates $1$ random link with other subpopulations. We then tackle the case of a general value of $m$.

\subsubsection{Case $m=1$}

Let us consider a SIR epidemic process started from a single node,  see Figure~\ref{sk_activity} for a schematic representation. 
We define $D_a^n$ as the number of diseased nodes in the activity class $a$ and generation $n$. As seen in the previous sections, when the number of diseased subpopulations is small, it is possible and convenient to adopt a tree-like approximation relating $D_a^n$ with $D_a^{n-1}$. At each time step each node might be infected by other seeded subpopulations if:
\begin{itemize}
\item  The node is active and connects with a seeded subpopulation that successfully infects it. 

Given $\lambda_{a'a}$ seeds arriving from a nodes in the class $a'$ to the nodes of class $a$, the destination will experience an outbreak with probability $1-R_0^{-\lambda_{a'a}}$.
\item The node is not active but is selected by an active and seeded subpopulation that seeds it.
\end{itemize} 
Hence, there are two different contributions:

\begin{eqnarray}
\label{eq_m1}
D_a^n &=& a \Delta t V_a\sum_{a'}D_{a'}^{n-1}\left( 1-R_0^{-\lambda_{a'a}} \right)\left( 1-\frac{D_{a}^{n-1}}{V_a}\right)\frac{1}{V}+ \nonumber \\
&+&  V_a\sum_{a'}a' \Delta t D_{a'}^{n-1}\left( 1-R_0^{-\lambda_{a'a}} \right)\left( 1-\frac{D_{a}^{n-1}}{V_a}\right)\frac{1}{V}. \nonumber \\
\end{eqnarray}

\noindent The value $\lambda_{a'a}$ represent the average number of seeds arriving from a node of class $a'$ to a node of class $a$. In analogy with Eq.~\ref{ll} we can write:

\be
\lambda_{a'a}=d_{a'a}\frac{\alpha}{\mu}N_{a'} \sim p \Delta t \frac{\alpha}{\mu}N_{a'}.
\ee

\noindent  Indeed, for $m=1$ the degree per unit time of nodes that can infect or can be infected is $1+\av{a}$. They have at least one connection, plus a contribution given by the average value of activity of all the other nodes. For power-law distributions of activity $\av{a}\sim \mathcal{O}(\epsilon)\ll1$, and we can approximate the degree to $1$.

We make further approximations and simplification by setting without losing generality $\Delta t=1$, and by considering (i) that for values of $R_0$ close to one $( 1-R_0^{-\lambda_{a'a}}) \sim \lambda_{a'a}(R_0-1)$, and (ii) that in the early stages of the spread the large majority of nodes are not seeded so that $( 1-\frac{D_{a}^{n-1}}{V_a})\sim 1$. We can write:
\begin{eqnarray}
\label{eq_m1_2}
D_a^n &=& a F(a)\Omega \sum_{a'}D_{a'}^{n-1}N_{a'}+ \nonumber \\
&+& F(a)\Omega\sum_{a'}a'D_{a'}^{n-1}N_{a'}, \nonumber \\
\end{eqnarray}
where $\Omega=p(R_0-1)\frac{\alpha}{\mu}$. In general, it is not easy to solve directly this equation. In order to find a close form we therefore introduce a set quantities:
\begin{eqnarray}
\theta^n &=& \sum_a D_a^n N_a\nonumber \\ 
\xi^n &=& \sum_a a D_a^n N_a  \nonumber \\
\phi_h &=& \sum_a a^{h}F(a)N_a, \;\ h=1,2 \nonumber \\
\end{eqnarray}
By multiplying both sides of  Eq.~(\ref{eq_m1_2}) by $N_a$ and summing, we obtain
\be
\label{eq22}
\theta^n =\phi_1 \Omega \theta^{n-1}+\Omega \av{N} \xi^{n-1}
\ee
On the other hand, by multiplying both sides of  Eq.~(\ref{eq_m1_2}) by $aN_a$ and summing, we obtain
\be
\label{eq33}
\xi^n=\phi_2\Omega\theta^{n-1}+\Omega \phi_1\xi^{n-1}
\ee

In the continuous time limit, we can write Eqs.~(\ref{eq22}) and~(\ref{eq33}) in a differential form:
\begin{eqnarray}
 \partial_{n} \theta &=&(\phi_1\Omega-1)\theta^{n-1}+\Omega \av{N}\xi^{n-1}\\
\partial_{n} \xi &=&\phi_2\Omega\theta^{n-1}+(\phi_1\Omega-1)\xi^{n-1}
\end{eqnarray}

The Jacobian matrix of this set of linear differential equations takes the form
\[ J= \left( \begin{array}{cc}
\phi_1\Omega-1 & \Omega \av{N}  \\
\phi_2\Omega & \phi_1\Omega-1  \\ \end{array} \right)
\] 
and has eigenvalues
\be
\Lambda_{(1,2)} = \phi_1\Omega-1 \pm \Omega \sqrt{\phi_2 \av{N}}
\ee
The epidemic threshold is obtained by requiring the largest eigenvalues to be larger than $0$, which leads to the condition for the global outbreak:
\be
\Omega(\phi_1+\sqrt{\phi_2 \av{N}} >1
\ee

\noindent we can write the threshold condition on the mobility rate 

\be 
\lb{thre_1}
p^* = \frac{\mu }{(R_0-1)\alpha}\frac{1}{\phi_1+\sqrt{\phi_2 \av{N}}}.
\ee

This equation is the central result of our analysis, and it is worth highlighting some aspects of it. First, the invasion threshold is not function of the time-integrated network representation. It depends on two functions $\phi_1$ and $\phi_2$. As shown in details before, the first quantity is the average number of individuals potentially moving out from active nodes. The second quantity is an higher order function of less clear physical interpretation. Second, both quantities act together in lowering the value of the threshold, meaning that not only the amount of mobility observed in the network is responsible for an easier outbreak of the contagion process, but also its heterogeneity act in the same way. Finally, Eq.~(\ref{thre_1}) shows that a thorough consideration of the time-varying nature of the network is crucial not to introduce biases in the description of the processes, also in metapopulation systems.

We test the analytical prediction of Eq.~(\ref{thre_1}) by extensive analytical simulations. Considering $V=10^5$, $m=1$, $F(a)=Aa^{-2.2}$, $\epsilon=10^{-3}$, $\mu=10^{-2}$ and $R_0=1.1$ we evaluated the number of diseased subpopulation as a function of $p$.  Figure~\ref{fig3}-A confirms that the analytical value perfectly matches the simulations. 
It also shows that at large values of $p$ the number of diseased subpopulations reaches a maximum, and decreases afterwards. This is in an interesting behavior due to the particular choice of $m=1$ in the activity driven model. When $p$ is large, a connected node loses all of its individuals immediately. Thus, at each time step very active nodes give to a single, and with high probability less active, node all the seeds, which then reside for a longer time on the new node. This turns out to limit strongly the number of possible new seeded subpopulations that can be generated. In Figure~\ref{fig3}-B we show the number of diseased subpopulation as a function of $p$, and $R_0$. We fixed  $V=10^4$, $m=1$, $F(a)=Aa^{-2.2}$, $\epsilon=10^{-3}$, and $\mu=10^{-2}$. The plot clearly show the validity of the analytical formulation for a wide range of reproductive numbers.

\begin{figure}
\begin{center} 
 \includegraphics  [width=0.4\textwidth] {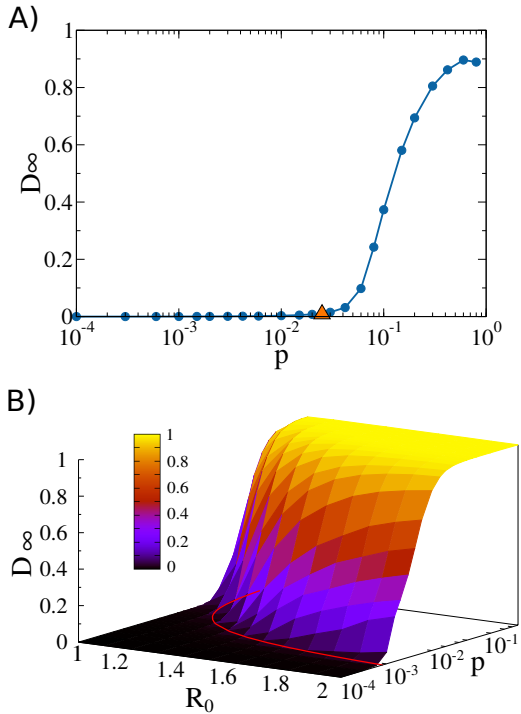}
\caption{(Color Online) In panel A) we show the number of diseased nodes in a metapopulation network, $D_\infty$, as a function of mobility rate $p$. We consider $V=10^5$ subpopulations, $\av{N}=10^3$ individuals per subpopulation, $m=1$, $\epsilon=10^{-3}$ and $\eta=1$. We select a power-law distribution of activity potential $F(x) \sim x^{-\gamma}$ with $\gamma=2.2$. We then consider $\mu =10^{-2}$, and $R_0=1.1$. The triangle represents the analytical prediction of the invasion threshold according to Eq.~\ref{thre_1}. In panel B) we show the phase space of a SIR process. Considering  $V=10^4$ subpopulations, $\av{N}=10^3$ individuals per subpopulation, $m=1$, $\epsilon=10^{-3}$ and $\eta=1$. We select a power-law distribution of activity potential $F(x) \sim x^{-\gamma}$ with $\gamma=2.2$. The 3D surface represents the value of the final number of diseased nodes in the metapopulation system as a function of the local threshold $R_0$ and of the diffusion rate $p$. The red  (light gray) curve defines the global diffusion threshold $p_
c$ for each $R_0$. Each plot is made averaging $10^2$ independent simulations. Each one of them started with $1$\% of random seeds.}
\lb{fig3}
\end{center}
\end{figure}

\subsubsection{General case, $m>1$}

The general case of an arbitrary number of links established by active nodes can be derived following the same approach described before. The only difference is the expression of $\lambda_{a'a}$ that for active nodes now becomes:
\be
\lambda_{a'a}\sim \frac{p}{m}\frac{\alpha}{\mu}N_a',
\ee
where we have neglected the contribution to the degree given by other active nodes.
The value of $\lambda_{a'a}$ for inactive nodes, that are selected by active ones, is the same as in the $m=1$. Thus, the process is now asymmetric. Active and passively selected nodes have in general a different degree and a different capacity to accumulate individuals. Eq~(\ref{eq_m1_2}) becomes:
\begin{eqnarray}
\label{eq_m1_3}
D_a^n &=& a F(a)m \Omega \sum_{a'}D_{a'}^{n-1}N_{a'}+ \nonumber \\
&+& F(a)\Omega\sum_{a'}a'D_{a'}^{n-1}N_{a'}. \nonumber \\
\end{eqnarray}
Following the same steps and introducing the same quantities we get a the threshold:
\be
\lb{thre_2} 
p^* = \frac{\mu }{(R_0-1)\alpha}\frac{2}{(m+1)\phi_1+\sqrt{4m\overline{N}\phi_2+(m-1)^2\phi_1^2 }}.
\ee

Remarkably, while this result reduces to the previous one, Eq.~(\ref{thre_1}), when $m=1$, the functional form is now different for a general value of $m$. This is due to the possibility of active nodes to release individuals to more than one single subpopulation, which interestingly does not translate in a simple shift of $p^*$. However, also in this case the critical value is a function of the number of individuals moving due to active and selected nodes, and of its heterogeneity, which again act together in determining a smaller threshold.

We test the analytical prediction of Eq.~(\ref{thre_2}) by extensive analytical simulations. Considering $V=10^5$, $m=3$, $F(a)=Aa^{-2.2}$, $\epsilon=10^{-3}$, $\mu=10^{-2}$ and $R_0=1.1$ we evaluated the number of diseased subpopulation as a function of $p$. As show in Figure~\ref{fig4}-A the analytical value perfectly matches the simulations. 
For values of $m>1$ the behavior observed for large $p$ and $m=1$  disappears as expected. In Figure~\ref{fig4}-B we show the number of diseased subpopulation as function of $p$, and $R_0$. We consider $V=10^4$, $m=3$, $F(a)=Aa^{-2.2}$, $\epsilon=10^{-3}$, and $\mu=10^{-2}$. Also in this case the plot confirms the analytical prediction for the mobility threshold.

\begin{figure}
\begin{center} 
 \includegraphics  [width=0.4\textwidth] {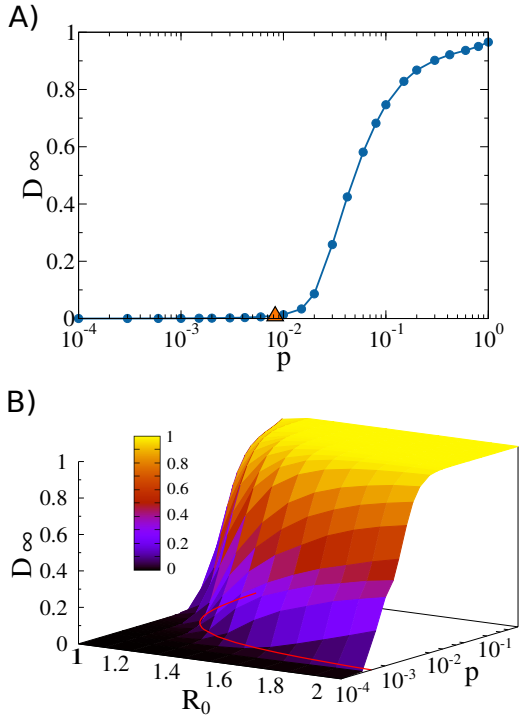}
\caption{(Color Online) In panel A) we show the number of diseased nodes in a metapopulation network, $D_\infty$, as a function of mobility rate $p$. We consider $V=10^5$ subpopulations, $\av{N}=10^3$ individuals per subpopulation, $m=3$, $\epsilon=10^{-3}$ and $\eta=1$. We select a power-law distribution of activity potential $F(x) \sim x^{-\gamma}$ with $\gamma=2.2$. We then consider $\mu =10^{-2}$ and $R_0=1.1$. The triangle represents the analytical prediction of the invasion threshold according to Eq.~\ref{thre_2}. In panel B) we show the phase space of a SIR process. Considering  $V=10^4$ subpopulations, $\av{N}=10^3$ individuals per subpopulation, $m=3$, $\epsilon=10^{-3}$ and $\eta=1$. We select a power-law distribution of activity potential $F(x) \sim x^{-\gamma}$ with $\gamma=2.2$. The 3D surface represents the value of the final number of diseased nodes in the metapopulation system as a function of the local threshold $R_0$ and of the diffusion rate $p$. The red (light gray) curve defines the global diffusion threshold $p_
c$ for each $R_0$. Each one of them started with $1$\% of random seeds.}
\lb{fig4}
\end{center}
\end{figure}

In order to measure more directly the effects induced by time-varying couplings, we compare the behavior of the fraction of diseased subpopulations, $D_\infty $, for activity driven networks integrated over $T$ time steps with the explicit dynamic counterpart we consider so far. In the integrated networks links are accumulated over time, and the spreading process is run in the resulting static graph i.e. $\mathcal{G}_{int}=\cup_{t=0}^{t=T}\mathcal{G}_t$. As show in Figure~\ref{both} when the explicit dynamics of the connections among subpopulation is neglected and integrated over time the critical value of the mobility threshold is orders of magnitudes smaller. This plot clear show the biases that might be introduced by time integrating representations of the system and the importance of considering dynamical coupling patterns for processes that evolve with comparable timescales. These results nicely generalize to metapopulation systems the findings obtained in networks constituted by the interactions of 
single individuals~\cite{perra12-1,perra12-2}.

\begin{figure}
\begin{center} 
 \includegraphics  [width=0.4\textwidth] {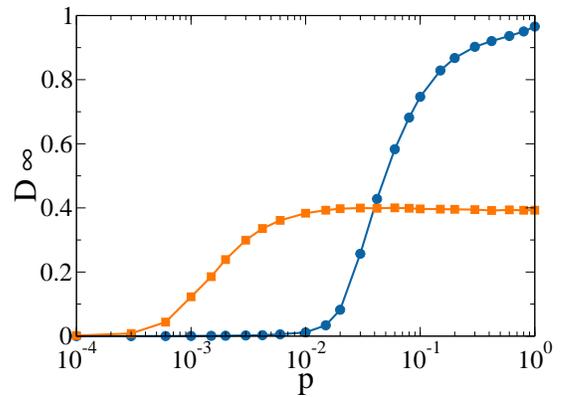}
\caption{(Color Online) The plot shows the number of diseased nodes in a metapopulation network, $D_\infty$, as a function of mobility rate $p$ on activity driven networks (dots) and corresponding integrated, static, ones (squares). The integration has been performed considering $50$ time steps. We fix $V=10^4$ subpopulations, $\av{N}=10^3$ individuals per subpopulation, $m=3$, $\epsilon=10^{-3}$ and $\eta=1$. We select a power-law distribution of activity potential $F(x) \sim x^{-\gamma}$ with $\gamma=2.2$. We then consider $\mu =10^{-2}$ and $R_0=1.1$. Each plot is made averaging $10^2$ independent simulations. Each one of them started with $1$\% of random seeds.}
\lb{both}
\end{center}
\end{figure}

\section{Conclusions}
\lb{conclu}

In this paper we have analyzed reaction-diffusion processes in metapopulation networks characterized by time varying coupling patterns, described in the activity driven framework. We have focused on the spreading and contagion processes of diseases and information, modeled through a SIR model. Here, the reaction dynamics inside each node is described considering a homogeneous mixing approach, where each pair of individuals has the same probability of interact, while the diffusion of individuals between connected nodes is Markovian. In this setting, we have characterized analytically the conditions for a global spreading as a function of the process parameters and the time varying couplings. Noteworthily, they are strikingly different from the static case, and turn out to be a function of the average number of individuals potentially moving out from active nodes, that in the early phases of the process spread the diseases to the selected target nodes. These findings not only provide the first analytical 
insight into time varying metapopulation processes, but clearly point out that dynamical couplings affect spreading phenomena also in reaction-diffusion processes, and that important biases might be introduced by neglecting them. 

Recent analysis have shown that real time-varying networks are characterized by strong dynamical correlations and memory~\cite{holme11-1}. These complex properties are difficult to describe analytically, and in this paper we have provided a first description of reaction-diffusion processes in the simpler setting of a Markovian, memoryless, time-varying metapopulation networks with no dynamical correlations. Furthermore, we have considered Markovian and homogenous random walks. Different mobility rules, as well as a full consideration of the dynamical properties of real-world networks, might introduce non-trivial effects on the global dynamics as observed for static metapopulation networks~\cite{colizza07-5}. All these features are definitely important and call for future extensions of the present framework. 
\section{Acknowledgments}

The authors are grateful to Alessandro Vespignani for helpful discussions, insights, and comments. The authors thank also an anonymous referee for helpful suggestions that helped improving the manuscript.

\end{document}